\documentstyle [12pt,twoside]{article}

\renewcommand{\theequation}{\arabic{section}.
\arabic{equation}}

\def\be{\begin{equation}}
\def\te{\end{equation}}

\def\bea{\begin{eqnarray}}
\def\nn{\nonumber}
\def\tea{\end{eqnarray}}

\def\leftrightarrowfill{$\mathsurround=0pt \mathord\leftarrow \mkern-6mu
        \cleaders\hbox{$\mkern-2mu \mathord- \mkern-2mu$}\hfill
        \mkern-6mu \mathord\rightarrow$}
\def\overleftrightarrow#1{\vbox{\ialign{##\crcr
        \leftrightarrowfill\crcr\noalign{\kern-1pt\nointerlineskip}
        $\hfil\displaystyle{#1}\hfil$\crcr}}}

\def\frac#1#2{{\textstyle{#1\over\vphantom2\smash{\raise.20ex
        \hbox{$\scriptstyle{#2}$}}}}}

\def\sfrac#1#2{{\vphantom1\smash{\lower.5ex\hbox{\small$#1$}}\over
        \vphantom1\smash{\raise.4ex\hbox{\small$#2$}}}}
\def\bfrac#1#2{{\vphantom1\smash{\lower.5ex\hbox{$#1$}}\over
        \vphantom1\smash{\raise.3ex\hbox{$#2$}}}}
\def\afrac#1#2{{\vphantom1\smash{\lower.5ex\hbox{$#1$}}\over#2}}

\catcode`@=11
\def\underline#1{\relax\ifmmode\@@underline#1\else
        $\@@underline{\hbox{#1}}$\relax\fi}
\catcode`@=12

\newskip\humongous \humongous=0pt plus 1000pt minus 1000pt
        \crcr#1\crcr}}

\def\title#1#2#3#4{
        {\hbox to\hsize{#1 \hfill\ \ \  }}\par
        \begin{center}\vskip.5in minus.1in {\Large\bf #3}\\[.5in minus.2in]{#4}
        \vskip1.4in minus1.2in {\bf ABSTRACT}\\[.1in]\end{center}
        \begin{quotation}\par}

\def\begintitle#1#2#3#4
        {\begin{titlepage}
         \centerline{#1 \hfill }
         \begin{center}\vglue .7i
         {\large\bf #3}\\[.7in]
         {\bf {#4}}\\[.2in]
         {\bf ABSTRACT}\\
         \end{center}
         \begin{quotation}}
\def\endtitle
         {\end{quotation}
          \newpage}

\begin{document}
\pagestyle{empty}
\title{\ \ }{000}{\bf A Nonlinear Model of a Quantum Minisuperspace
System with Back Reaction}
{{\bf Marcos Rosenbaum} \\
{\bf Michael P. Ryan, Jr.}\footnote{e-mail: ryan@unamvm1.bitnet}\\
and \\{\bf Sukanya Sinha}
\footnote{e-mail: sinha@unamvm1.bitnet} \\
 {\it Instituto de Ciencias Nucleares, Universidad Nacional Aut\'onoma
de M\'exico\\
Circuito Exterior, C.U, A-Postal 70-543, M\'exico D.F 04510, MEXICO}}

We consider the quantum evolution of the space-independent mode of
a $\lambda {\phi}^4$ theory as a minisuperspace in the space of
all $\phi$. The motion of the wave packet in the minisuperspace is
compared to the motion of a wave packet in a larger minisuperspace
consisting of the original minisuperspace plus one space-dependent
mode. By comparing the motion of the two packets we develop
criteria that tell us when the quantum evolution in the
space-independent minisuperspace  gives us useful information about
the true evolution in the larger minisuperspace. These criteria
serve as a toy model for similar (but much more complex) criteria
that will tell us whether or when quantized gravitational
minisuperspaces can possibly give any useful information about
quantum gravity.

\endtitle

\setcounter{section}{0}
\section { Introduction}
\setcounter{equation}{0}
\pagestyle{plain}
\setcounter{page}{1}
For the past decade there has been considerable renewed interest in the
relatively old problem of quantum cosmology\cite {bib}.
The idea behind this field has been to
insert cosmological models directly into the ADM form of the
Einstein-Hilbert action. One then uses the Hamiltonian constraint thus
generated to construct a Wheeler-DeWitt equation that depends only on
the limited number of variables that define the cosmological model.
Finally this equation is used to find wave functions that hopefully
describe the quantum behavior of the universe
\cite{bib,Misner,DeWitt,Ryan}.
There are a number of
stumbling blocks to this program.  One is that plugging metrics with
a high degree of symmetry directly into the Einstein-Hilbert action
and varying that action with respect to the limited variables in the
action may not reproduce the classical Einstein equations for the
metric in question\cite {GRR1}.  We will not attempt to discuss this problem
here, but  will concentrate on a second problem.  This problem is the
fundamental one of whether the reduction of the action by symmetry
and its subsequent quantization produces a quantum system that means
anything physically.  Imposing high degrees of symmetry on a metric
and then quantizing implies putting metric variables
and their corresponding canonical momenta simultaneously equal to zero
and insisting that they remain identically zero for all time.  This
procedure is a direct violation of the uncertainty principle, and
cannot give a true quantum description of the problem.  The
conjecture, then, must be that somehow quantum cosmologies are an
approximation to some true solution of a full-fledged quantum
gravity.  Since there is as yet no such quantum theory of
gravity, we must attempt by various
subterfuges to construct at least plausibility arguments to show
whether there is any chance that quantum cosmologies contain any
physics.

A number of attempts have been made in this direction by
various groups.  Kucha\v r and Ryan have looked at both the $\lambda
\phi^4$ theory \cite{Kuchar} in flat space
and the Taub model imbedded in a general Bianchi
type IX model\cite{Kuchar2}, while Sinha and Hu \cite{Sinhu}
have looked at a $\lambda \phi^4
$ theory in a closed Robertson Walker background.
In all of these cases the idea was to take a
minisuperspace of a very high degree of symmetry  (a
``microsuperspace'') where the quantum problem is soluble, and imbed
it in a larger, but still simplified, field space (a
``minisuperspace'') that was still soluble (at the very least
approximately) in the quantum regime and compare the two solutions to
see if there is any way in which the microsuperspace solution can be
enough of an approximation to the full minisuperspace solution to
give reasonable physical information about the system.  Of course,
one must define exactly what type of approximation one expects the
quantum microsuperspace to be to the minisuperspace, and this is
where the different approaches diverge.  In Ref.\cite{Kuchar}
  the definition
was based on the idea of starting with the full Schr\"odinger
equation for superspace (the minisuperspace in our language),
expanding this wave function in eigenstates of the minisuperspace
Hamiltonian that parametrically depend on the microsuperspace
variables , and finding the conditions under which the full
Schr\"odinger equation could be reduced to a `` projected
Schr\"odinger equation " for the ``microsuperspace wave functions"
(the coefficients of the mode expansion that depend on the
microsuperspace variables) evolved by the microsuperspace
Hamiltonian. The expectation values for dynamical variables
on microsuperspace can then be calculated using a density matrix
constructed from the wave functions.
The approach in Ref.\cite{Sinhu} was very similar in spirit to
that of Ref.\cite{Kuchar}. The idea was to derive an ``effective"
Wheeler-
DeWitt equation for the microsuperspace sector
where the effect of the higher modes appeared as a
backreaction term, and the microsuperspace description was considered
good when the backreaction was small compared to the microsuperspace
potential. It was also shown that the backreaction could be
physically interpreted as a dissipative term arising from particles
produced in the ambient superspace modes due to the dynamical
evolution of the microsuperspace degrees of freedom and hence the
above criterion could be interpreted as a requirement of a low
rate of particle production.
In Ref.\cite{Kuchar2}
the approach was to investigate the behavior of a wave packet in the
minisuperspace strongly peaked initially around the microsuperspace
sector, and compare this behavior with that of a wave packet in the
microsuperspace.

The common denominator of all of these approaches can best be
visualized by noticing that a minisuperspace is formed by setting most
of the infinite number of parameters that describe a general
gravitational field equal to zero, often (but not always) leaving
only a finite number.  That means that a microsuperspace is a
hyperplane in the whole of superspace, and a quantum minisuperspace
solution is a solution constrained to lie in this hyperplane.  One
may think of the full superspace that surrounds the hyperplane as an
``environment'' of the minisuperspace, and because of the uncertainty
principle the system, no matter how confined it may be initially to
the minisuperspace hyperplane, must ``feel'' the influence of the
larger superspace.  This influence of the larger superspace may or
may not change the behavior predicted by the minisuperspace
quantization.  A minisuperspace may be considered ``good'' or ``bad''
depending on whether the influence of the environment changes the
physical predictions made using the minisuperspace sufficiently to
make them inviable.  The approaches discussed above differ mainly in
the type of minisuperspace predictions that one considers important, and
consequently on the way the environment affects them.

One cannot expect that all minisuperspaces will be ``good'', since
the environment that surrounds them will depend on where the
minisuperspace hyperplane lies in the full superspace.  In
Ref.\cite{Kuchar}
the idea of finding criteria that would tell us whether we have any
right to expect that a particular minisuperspace is ``good'' or
``bad'' was discussed.  In all of the gravitational examples found so
far, the Wheeler-DeWitt equation
$$G_{ijk\ell} {{\delta^2 \Psi}\over {\delta g_{ij} \delta
g_{k\ell}}} + gR\Psi = 0 \eqno(1.1)$$
has been used to quantize the system, and while the behavior of the
supermetric $G_{ijk\ell}$ near the minisuperspace hyperplane should
affect the full superspace quantization that is supposed to be
approximated by the minisuperspace quantization,
in all the gravitational cases considered so far the superspace
has been flat, so the major influence
has been the behavior of the ``potential'' term $gR$. Nevertheless,
it is probably reasonable to expect that in many (if not most) cases
$gR$ will dominate the quantum behavior of the gravitational system.

Before looking at as complicated a problem as the behavior of $gR$
near a minisuperspace, we felt that it might be worthwhile
to investigate a simpler minisuperspace problem and develop criteria
in such a context that would model the more complicated
criteria that one would expect
to find in quantum geometrodynamics.   We have chosen a
one-plus-one $\lambda
\phi^4$ scalar field theory similar to that studied in
Ref.\cite{Kuchar} , but
we will analyze it using an approach closer to \cite{Kuchar2}
, so that
we get a more complete understanding of this model from different
points of view before progressing to gravity.  The
action for our field is
$$S = \int [{{1}\over {2}} \dot \phi^2 - {{1}\over {2}}(\partial_z
\phi )^2 - {{\mu^2}\over {2}} \phi^2 - \lambda \phi^4 ]dz dt. \eqno
(1.2)$$
We will assume an $S^1$ topology for $t$ = const. slices, identifying
the end points $z = \pm L/2$.  This means that it is
possible to express $\phi$ in
terms of a countable number of modes in the form
$$ \phi = \phi_0 (t) + \sum_{n = 1}^{\infty}{\phi_n (t) \cos ({{2\pi
n}\over {L}} z) + \phi^{(s)}_n (t) \sin ({{2\pi n}\over {L}} z)}.
\eqno (1.3)$$
The ``minisuperspace'' of our problem will be the space-independent
mode $\phi_0 (t)$, and we will quantize this system after putting all
of the $\phi_n = 0$.  We will investigate the region of superspace
near the minisuperspace by means of the techniques used by Halliwell
and Hawking\cite{H-Hawk} to study regions of superspace near the $k = +1$
FRW cosmology, that is, we will substitute (1.3) into the action (1.2) and
keep terms to second order in the $\phi_n$.  This procedure will give
an action that when varied with respect to $\phi_n$ will give a
linear equation for $\phi_n$ that contains $\phi_0$ , and when varied
with respect to $\phi_0$ a linear equation for $\phi_0$ that
includes a back-reaction term of the $\phi_n$ on the $\phi_0$.  This
approximation will be discussed in more detail in Sec. 2.

Even at the second-order level in $\phi_n$, there are problems of
renormalization (albeit relatively simple ones), and to avoid these
we will use the idea of Ref.\cite{Kuchar} of putting all of the $\phi_n = 0$
(``freezing'' them) except for one (which is ``unfrozen'').  This is
then a minisuperspace itself, and we will call the $\phi_0 (t)$ mode
a ``microsuperspace'' imbedded in the two-dimensional minisuperspace
$(\phi_0, \phi_n )$. We differ from Ref.\cite{Kuchar}  in that we take
$\phi_n$ to be any of the $\phi_n$ of (1.3), whereas in
Ref.\cite{Kuchar}
only the lowest  space-frequency mode $\phi_1$ was taken. For the
approximation we will study, it can be shown that the choice of
$\phi_n$, $n > 1$ gives results which resemble those obtained by
studying the full field theory with
a cutoff at some frequency larger than $n\pi\over L$, and since $n$
is unspecified, we can use our system to model a cutoff at some
high frequency.

This toy model allows us to study the question of whether the
microsuperspace $\phi_0$ is a good microsuperspace in the
$(\phi_0 , \phi_n )$ minisuperspace.  As in the relativity cases
studied so far, we expect the potential term, $\mu^2 \phi^2 +
\lambda \phi^4$, in the region of the microsuperspace will be the
determining factor in deciding whether $\phi_0$ is a ``good''
microsuperspace or not.  The only free parameters here are $\mu^2$,
$\lambda$ and $L$, the size of the $t$ = const. slices.  In order to
leave room for several possible cases, we will not assume that either
$\mu^2$ or $\lambda$ is necessarily positive.  The only other free
parameter in the system is $n$, the mode number of the mode $\phi_n$.
The criteria we will develop for the usefulness of the
microsuperspace quantization will depend on the sizes of $\mu^2$,
$\lambda$, $L$, and $n$, and the signs of $\mu^2$ and $\lambda$.

We will base our criteria on the motion of wave packets, one in the
microsuperspace and one in the minisuperspace centered around the
microsuperspace.
In order to do this we will need wave packets
that are as close as we can find to coherent states for the system.
In Appendices A and B we develop what seems to us to be a new approach to
finding such states and find approximate solutions to the equations
generated by the approach.  Given these solutions we can compare the
solutions of the microsuperspace and minisuperspace quantum systems
and decide whether the fact that the minisuperspace packet ``feels''
the superspace surrounding the microsuperspace changes its behavior
sufficiently from that of the pure microsuperspace packet to
invalidate the microsuperspace approximation.

There is one other major problem about defining a ``good''
minisuperspace in the wave packet scheme.  The existence of a
superspace wave packet that stays near the minisuperspace packet may
not be enough to classify a minisuperspace as ``good''.  In Ref.
\cite{GRR2}
an example was given in the cosmological context of a minisuperspace
wave packet that remained centered around a microsuperspace, but was
unstable against small changes in initial conditions.  Whether such
instability should be taken to show that a minisuperspace is ``bad''
is debatable.  We will discuss this idea in the context of our model
theory in Sec. 4.

The structure of the article is as follows.  In Sec. 2 we will
discuss the classical action of the system we want to study and
write down the classical and quantum equations of motion.  Section
3 will cover the approximate coherent states we will use to try to
set up criteria for reasonableness of microsuperspaces, the details
of the derivation of which are given in Appendices A and B.  In
Sec. 4 we will develop and discuss the criteria.  Finally, in Sec. 5
we will analyze these criteria and try to relate them to the as yet
unknown criteria that might be expected for geometrodynamic
minisuperspaces, and then include some suggestions for further research.
\section{ Perturbations and their Back-reaction on the
Microsuperspace Sector}
\setcounter{equation}{0}
As we mentioned in the Introduction, the one-plus-one $\lambda \phi^4$ field
theory has minisuperspace sectors just as gravitation does.  The
minisuperspace closest in concept to the cosmological minisuperspace
is the space-independent mode of $\phi$, $\phi_0 (t)$.  This
minisuperspace has the property that if it is plugged into the
action (1.2)
it gives a reduced action
\be
S = L\int [{{1}\over {2}} \dot \phi_0^2 - {{\mu^2}\over {2}}
\phi_0^2 - \lambda \phi_0^4 ]dt,
\te
where we have assumed that the $t$ = const. slices have an $S^1$ topology
with $z = \pm
L/2$ identified.  When the action is varied with respect to $\phi_0$
it gives the correct equation of motion for $\phi_0$,
\be
 \ddot \phi _0 + \mu^2 \phi_0 + 4\lambda \phi_0^3 = 0.
\te
Of course, the Hamiltonian form of (2.2),
\be
 S = L\int [\pi_0 \dot \phi_0 - ({{1}\over {2}} \pi^2_0 +
{{\mu^2}\over {2}} \phi_0^2 + \lambda \phi_0^4 )] dt
\te
also gives the correct equations of motion for $\pi_0$ and $\phi_0$.

As we also mentioned in the Introduction, minisuperspace quantization of
the $\phi_0$ mode consists of quantizing the theory given by
(2.3) with $\phi_0$ and $\pi_0$, the configuration variable and its
conjugate momentum as the only variables of the problem, and this
quantization can be taken as a simplified toy model of quantum
cosmology.  We will not attempt at this point to quantize this system
but will give an approximate solution as part of the more general
problem discussed below.

As a model for studying when the quantization of a gravitational
quantum minisuperspace gives useful information, we will extend our
$\lambda \phi^4$ minisuperspace in a way similar to that used by
Halliwell and Hawking\cite{H-Hawk} to study the volume of superspace near the
$k = +1$ FRW minisuperspace.  They used a truncated action which in
our case can be constructed by first expanding the full $\phi (z,t)$
in a real Fourier series of the form of (1.3),
where for convenience we will set all of the $\phi^{(s)}_n (t)$ equal
to zero since the final result will not be qualitatively different if
these terms are kept.  If we put this form of $\phi$ into (1.2) we find
\bea
S &=& L \int [ {{1}\over {2}} \dot \phi_0^2 + {{1}\over {4}} \sum_n
\dot \phi_n^2 - {{\mu^2}\over {2}} \phi_0^2 - \lambda \phi_0^4 -
{{1}\over {4}} \sum_n \phi_n^2 \left ({{2\pi n}\over {L}} \right )^2
- {{\mu^2}\over {4}} \sum_n \phi_n^2 -\nn \\
& &-
\lambda \phi_0 \sum_{n,m,\ell} A_{nm\ell} \phi_n \phi_m \phi_{\ell}
- 3\lambda \phi_0^2 \sum_n \phi_n^2 - \lambda \sum_{n,m,\ell ,k}
B_{mn\ell k} \phi_n \phi_m \phi_{\ell} \phi_k ]dt,
\tea
where $A_{mn\ell}$ and $B_{mn\ell k}$ are constants given by the
integrations over products of $\cos (2n\pi z/L)$.  The truncated
action assumes that the $\phi_n$ are small enough that the cubic and
quartic terms in the $\phi_n$ are negligible, while the quadratic
terms are large enough to affect the behavior of the $\phi_0$. In
this case the action reduces to
\be
 S = L \int [{{1}\over {2}} \dot \phi_0^2 + {{1}\over {4}}\sum_n
\dot \phi_n^2 - {{\mu^2}\over {2}} \phi_0^2 - \lambda \phi_0^4 -
{{\mu^2}\over {4}} \sum_n \phi_n^2 - {{1}\over {4}} \sum_n \phi_n^2 \left
({{2\pi n}\over {L}} \right )^2 - 3 \lambda \phi_0^2 \sum_n \phi_n^2
]dt.
\te
We will not attempt here to justify the exclusion of the quartic and
cubic terms, but will instead discuss the meaning of the classical
equations of motion derived from (2.5).  The equation for $\phi_0$ is
\be
\ddot \phi_0 + \mu^2 \phi_0 + 4\lambda \phi_0^3 + 6\lambda \phi_0
\sum_n \phi_n^2 = 0,
\te
while the equations for the $\phi_n$ are
\be
\ddot \phi_n + \mu^2 \phi_n + \left ({{2\pi n}\over {L}} \right
)^2 \phi_n + 12 \lambda \phi_0^2 \phi_n = 0.
\te
Up to the order we have kept the $\phi_n$ these are exact.
These exact equations can be interpreted as simple perturbation
equations for the $\phi_n$ and (2.6) as an equation for $\phi_0$ that
includes the first non-zero term of the back reaction of the $\phi_n$
on the $\phi_0$.  However, this interpretation only makes sense if we
assume that the amplitudes $\phi_n$ are more or less randomly
distributed, and are not correlated in such a way as to represent a
large concentration of field $\phi$ at some point $z = z_0$.  Perhaps
the best way to qualify this idea is to call it a ``cosmological''
paradigm.  The usual picture of the universe is that it is made up of
matter condensations that are locally inhomogeneous, but that these
condensations are spread throughout space in such a way that on the
average the matter density is homogeneous.  It is this averaged
density that drives the spatially homogeneous gravitational mode.  Of
course, any gravitational field can be broken up into harmonic modes,
and the homogeneous mode will be affected by the inhomogeneous modes,
but if, say, the universe is half empty, the homogeneous mode cannot
reasonably be interpreted as a homogeneous cosmological background
driven by the averaged matter density.

In order to study the meaning of quantum minisuperspace solutions in
the context of the present theory, we want to first solve the quantum
problem in the microsuperspace sector where the $\phi_n$ are set
equal to zero, and in this case the action is (2.3) and corresponds
to a one-dimensional anharmonic oscillator.  We are not assuming that
either $\mu^2$ or $\lambda$ is necessarily positive, so there are four
possible microsuperspace quantum systems, Case I ($\mu^2 >0, \lambda >
0$), Case II ($\mu^2 < 0, \lambda > 0$), Case III ($\mu^2 > 0, \lambda <
0$), and Case IV ($\mu^2 < 0, \lambda < 0$).  We can quantize the
problem given by (2.3) by realizing the operator $\hat \pi_0$ as $-i
\partial / \partial \phi_0$ operating on a state function $\Psi
(\phi_0, t)$, which leads to the Schr\"odinger equation
\be
-{{1}\over {2}} {{\partial^2 \Psi}\over {\partial \phi^2_0}} +
{{\mu^2}\over {2}} \phi_0^2 \Psi + \lambda \phi_0^4 \Psi = i
{{\partial \Psi}\over {\partial t}}.
\te
We would like to compare the solutions of this equation with
solutions for the full superspace of all possible modes, ($\phi_0,
\phi_n$) in order to see if the presence of the higher modes,
$\phi_n$, affects the behavior of the $\phi_0$ mode in such a way as
to cause drastic changes in its quantum dynamics.  However,
as we said in the Introduction, in order
to avoid renormalization problems we will work with a simpler system
similar to that used by Kucha\v r and Ryan \cite{Kuchar}, where we will imbed
the $\phi_0$ (microsuperspace) mode in an extended minisuperspace
where only one of the $\phi_n$ is non-zero .
Since in the truncated action the $\phi_n$ modes do not interact
among themselves, this ansatz is consistent.

We will now need the Hamiltonian form of (2.5).
In order to simplify our notation, we will define $x \equiv \sqrt{L}
\phi_0$, $y \equiv \sqrt{L/2}\phi_n$, and $p_x$ and $p_y$ their
conjugate momenta.  With these definitions the
final action for $x$ and $y$ becomes
\be
 S = \int [p_x \dot x + p_y \dot y - \{ {{1}\over {2}}p_x^2 +
{{1}\over {2}}p_y^2 + {{\mu^2}\over {2}} x^2 + {{\lambda}\over {L}}
x^4 + [{{\mu^2}\over {2}} +
{{1}\over {2}}\left({{2\pi n}\over {L}} \right)^2]y^2 +
6{{\lambda}\over {L}}x^2 y^2 \}]dt.
\te
In this case the minisuperspace state function $\Psi (x, y, t)$ obeys
the Schr\"odinger equation
\be
-{{1}\over {2}} {{\partial^2 \Psi}\over {\partial x^2}} - {{1}\over {2}}
{{\partial^2 \Psi}\over {\partial y^2}} + {{\mu^2}\over {2}} x^2
\Psi + {{\lambda}\over {L}} x^4 \Psi +{{1}\over {2}}\left [ \mu^2 +
\left ( {{2\pi n}\over {L}} \right )^2 \right ] y^2 \Psi +
6{{\lambda}\over {L}} x^2 y^2 \Psi = i {{\partial \Psi}\over
{\partial t}}.
\te
Of course, the interaction term $6(\lambda /L) x^2 y^2 \Psi$ gives
the only influence of the $\phi_n$ mode on the $\phi_0$ mode.
To simplify our notation further, we will define $m^2 \equiv \mu^2
+ (2n\pi /L )^2$.  In the
next section we will find approximate solutions to (2.10), first a
microsuperspace solution where $y$ and $\partial^2 \Psi / \partial
y^2$ are put equal to zero and second a minisuperspace solution to the full
equation. The solution to the full (2.10) will be made as close as
possible to a coherent state centered around $y = 0$, and we will
investigate its behavior relative to the microsuperspace solution.
\def\Sm{\tilde {\cal S}}
\section{Quantum
Solutions for the Microsuperspace and Minisuperspace Models}
\setcounter{equation}{0}
In this section we will first write down the solution to the Schr\"odinger
equation for the microsuperspace model and subsequently for the full
minisuperspace model for Case I, i.e, ${\mu}^2 >0, \lambda >0$.
Since the quantum problem for even the
microsuperspace sector is not exactly solvable, we will
use ${\lambda \over L}$ as a perturbation parameter and obtain a perturbative
solution to first order in ${\lambda \over L}$.
However, since ultimately we are interested
in comparing the quantum dynamics of the truncated model with that of
the full model, we will look for time dependent
solutions that are analogous to
coherent states for these models,
rather than the stationary solutions that are
found in the usual applications of quantum perturbation theory .

Let us start by considering the Schr\"odinger equation for the
microsuperspace model. This can be written down from (2.8):
\be
i {\partial\Psi\over \partial t}
= {-{1\over 2}}{{\partial}^2\Psi\over {\partial x^2}}
+ \left({1\over 2}\mu^2 x^2 + {\lambda\over L} x^4 \right)\Psi .
\te
We make the following ansatz for the wave function
\be
{\Psi}_{micro} = e^{-{\cal S}}  ,
\te
where ${\cal S} = {\cal S}_0 + {\lambda\over L } {\cal S}_1$.
We will now substitute the above
ansatz in the Schr\"odinger equation (3.1) and
retain terms only up to linear order in ${\lambda \over L}$.  As far as we
know, this is the first attempt to apply perturbation theory to the
exponent ${\cal S}$ of $\Psi$.  Other attempts have directly
perturbed the function $\Psi$, and correspond to expanding
$\exp(-{\lambda\over L} {\cal S}_1)$ as $1 - ({\lambda\over L}) {\cal S}_1$.

Now, to lowest order we
obtain an equation for ${\cal S}_0$ given by,
\be
-i {\partial {\cal S}_0\over \partial t}
= -{1\over 2}{\left({\partial {\cal S
}_0\over \partial x }\right)}^2
{+{1\over 2}}{{\partial}^2 {\cal S}_0\over {\partial x^2}}
+ {1\over 2}\mu^2 x^2     ,
\te
and to $O({\lambda \over L})$ we obtain
\be
-i {\partial {\cal S}_1\over \partial t}
= - \left({\partial {\cal S}_0\over \partial x }\right)
\left({\partial {\cal S}_1\over \partial x }\right)
{+{1\over 2}}{{\partial}^2 {\cal S}_1\over {\partial x^2}}
+ x^4  .
\te
Let us concentrate on the zeroth order solution, i.e, ${\cal S}_0$ first. We
should mention that since this essentially only involves the quantum
solution to the harmonic oscillator problem, it is exactly solvable,
and the solutions are well known. In particular, the coherent state
solution is known, and we can write it down directly. However, we will
go through it in some detail merely to illustrate our technique of
obtaining it.
Separating ${\cal S}_0$ into real and imaginary parts as
${\cal S}_0 = {{\cal S}_0}^R + i{{\cal S}_0}^I$
, the problem now reduces to solving the real and imaginary parts of
eqn. (3.3 ), which are given by
\be
- {\partial{{\cal S}_0}^R\over \partial t}
= - \left({\partial {{\cal S}_0}^R\over \partial x }\right)
\left({\partial{{\cal S}_0}^I\over \partial x }\right)
+{1\over 2}{{\partial}^2 {{\cal S}_0}^I\over {\partial x^2}}
\te
and
\be
 {\partial{{\cal S}_0}^I\over \partial t}
= -{1\over 2}{\left({\partial {{\cal S}_0}^R\over \partial x }\right)}^2
+{1\over 2}{\left({\partial {{\cal S}_0}^I\over \partial x }\right)}^2
  {+{1\over 2}}{{\partial}^2 {{\cal S}_0}^R\over {\partial x^2}}
+ {1\over 2}\mu^2 x^2
\te
respectively.
Let us now specialize our ansatz to the following form ( clearly
suggested by the known coherent state solution):
\be
{\cal S}_0 = {\mu\over 2}{\left[x - g(t)\right]}^2 + iP(t)x ,
\te
where $P$ and $g$ are real functions of time. Substituting the
real and imaginary parts of ${\cal S
}_0$ from (3.7) into eqns. (3.5) and
(3.6)
respectively, we obtain the equations satisfied by these functions
by equating equal powers of $x$.
These are given by:
\be
\ddot{g}(t) + {\mu}^2 g(t) = 0
\te
and
\be
P(t) = -\dot{g} .
\te
Eqn. (3.8) is simply the equation for a classical harmonic oscillator
with coordinate $g$ and the solution we choose is
\be
g(t) = x_0 \cos \mu t ,
\te
where $x_0$ is a real constant determined by initial conditions. Then
$P(t)$ can be determined through eqn. (3.9 ) easily. We will not write
down the solution for $P(t)$ explicitly, since it corresponds to a
phase in the wave function, and we are ultimately interested in the
probability density $|\Psi(x)|^2$. Therefore, what we are after is
really the quantity $e^{-2{\cal S}^R}$, and as a result we will avoid
calculating the imaginary parts of ${\cal S}$ explicitly throughout the
rest of this paper. It is evident from eqn. (3.9) and (3.10) that
we have recovered the usual coherent state solution for the
wavefunction to the lowest order, and it is a Gaussian peaked around
the classical trajectory given by (3.10).  The rest of the solution
will represent a perturbation on this coherent state.

Thus
\be
{{\cal S}_0}^R = {\mu\over 2}{\left(x - x_0 \cos \mu t\right)}^2  .
\te

Let  us continue on to the $O({\lambda \over L})$ part.
The equations for the real and imaginary parts are given by
\be
- {\partial{{\cal S}_1}^R\over \partial t}
= - \left({\partial {{\cal S}_0}^R\over \partial x }\right)
\left({\partial{{\cal S}_1}^I\over \partial x }\right)
- \left({\partial {{\cal S}_0}^I\over \partial x }\right)
\left({\partial{{\cal S}_1}^R\over \partial x }\right)
+{1\over 2}{{\partial}^2 {{\cal S}_1}^I\over {\partial x^2}}
\te
and
\be
 {\partial{{\cal S}_1}^I\over \partial t}
= - \left({\partial {{\cal S}_0}^R\over \partial x }\right)
\left({\partial{{\cal S}_1}^R\over \partial x }\right)
+ \left({\partial {{\cal S}_0}^I\over \partial x }\right)
\left({\partial{{\cal S}_1}^I\over \partial x }\right)
+{1\over 2}{{\partial}^2 {{\cal S}_1}^R\over {\partial x^2}} + x^4 .
\te
Guessing from the form of the lowest order solution, we specialize
our ansatz further to
\bea
{{\cal S}_1}^R &=& \alpha x^4 + \beta x^3 + \gamma x^2 + \sigma x +
\omega ,\nn \\
{{\cal S}_1}^I &=& B x^3 + C x^2 + D x + E .
\tea
We follow exactly the same strategy as before to solve for the
unknown functions. We first substitute the ansatz (3.14) in eqns (3.12)
and (3.13) and then obtain equations satisfied by these functions by
equating equal powers of $x$. In this case of course, execution of
this strategy is much more complicated since we obtain coupled
equations for the functions. However, as shown as in  Appendix A,
these do indeed form a consistent set and can be successively
solved such that each satisfies a harmonic oscillator equation
driven by a source provided by a combination of the other functions
that have already been solved for.

Here we will  write down the final solution for only the real part
${{\cal S}_1}^R$. The coefficients $\alpha, \beta, \gamma, \sigma , \omega$
in (3.14) are given as follows
\bea
\alpha &=& {1\over 4\mu} ,\nn \\
\beta  &=& {x_1}^3\cos{3\mu t} + {x_0\over 2\mu}\cos{\mu t} ,\nn \\
\gamma &=& {3\over 4{\mu}^2} + C_1 \cos{2\mu t}
- {3\over 2}{x_0}^2{x_1}^3 \cos{4\mu t} + {3\over 2}{x_0}^2 t \sin
{2\mu t} ,\nn\\
\sigma &=& d_1 \cos {\mu t}
-{3\over 2\mu}\left[\left({1\over 2}{x_0}^3 + {x_1}^3
\right) - {x_0}C_1 \right]\cos {3\mu t} , \nn \\
& & + {3\over 4}{x_0}^2{x_1}^3
\cos {5\mu t}
 + {3x_0\over \mu}t \sin{\mu t} -
{3\over 2}{x_0}^3 t\sin{3 \mu t} ,\nn\\
\omega &=& -{1\over 4\mu}g^4 - {3\over 4{\mu}^2}g^2 - {3g\beta\over
2\mu}
 -g^3 \beta - g\sigma - {\gamma\over 2\mu} ,
\tea
where $x_1 , C_1 $ and $d_1$ are arbitrary constants.
As stated before, we do not need the imaginary part, but as shown in
the Appendix, since the equations are coupled we do need to solve
for $B, C,$ and $D$ to obtain all the coefficients of the real part.
Thus from equations (3.2) , (3.11) and (3.15) we have the complete
unnormalized wavefunction for the microsuperspace sector to
$O({\lambda \over L})$ in the exponent.

At this point we would like to comment on the so called secular
terms of the form $t\sin{2\mu t}$ appearing in the solution. Such
terms are known to appear in the straightforward application
of perturbation theory to solve the classical anharmonic oscillator
problem \cite{mscale}. In that context they are known to be pathological,
because it is known that the full solution must have a behavior
bounded in time and clearly the secular terms  have the
incorrect behavior since they grow with time. This turns out to
be an artifact of a naive application of perturbation theory, and the
correct perturbative solution without secular terms can be obtained
by using more sophisticated methods such as the method of multiple
scales \cite{mscale}. We believe that the ``quantum secular terms" in our
problem have the same origin, since the center of our almost
coherent state approximately obeys the classical equation of motion.
Thus we can probably get a solution free of secular terms by
an application the more sophisticated perturbation techniques adapted
to quantum mechanics  but we will not pursue this further in this paper.
We will therefore confine our analysis to short time scales before
the secular terms become dominant.

Now let us proceed to the full problem, i.e, to  solve the
Schr\"odinger equation for the minisuperspace model, corresponding to
the action given in eqn.(2.9) , where now we have an additional
coordinate $y$ that couples to the microsuperspace sector.
The Schr\"odinger equation in this case is given by the full eqn.
(2.10).
We follow exactly the same steps as before, with the ansatz
\be
{\Psi}_{mini} = e^{- \left({\Sm}_0 + {\lambda\over L} {\Sm}_1
\right)}.
\te
We notice that the lowest order problem is that of two uncoupled
harmonic oscillators with masses $\mu$ and $m$ corresponding to
the coordinates $x$ and $y$ respectively. So, we see from our
experience with the microsuperspace case that
$e^{-{\Sm_0}}$
can  be written down directly as
a product of two independent coherent state solutions for the
individual oscillators with ${\Sm}_0$ given by:
\be
{\Sm}_0  = {\mu\over 2}{(x - x_0 \cos {\mu t})}^2
+  {m\over 2}{(y - y_0 \cos {m t})}^2  + iPx + i\bar{P}y ,
\te
where $y_0$ is a constant to be fixed by initial conditions and
$\bar{P}$ is a known function of time satisfying an equation
identical to (3.9) in the $y$ coordinate.

The real and imaginary parts of the $O({\lambda \over L})$ equation obtained by
substituting the ansatz in the Schr\"odinger equation are given as
follows
\bea
 {\partial{{\Sm}_1}^I\over \partial t}
&=& - \mu (x - g)
\left({\partial{{\Sm}_1}^R\over \partial x }\right)
- \dot{g}
\left({\partial{{\Sm}_1}^I\over \partial x }\right)
+{1\over 2}{{\partial}^2 {{\Sm}_1}^R\over {\partial x^2}} + x^4 \nn \\
& & - m (y - f)
\left({\partial{{\Sm}_1}^R\over \partial y }\right)
- \dot{f}
\left({\partial{{\Sm}_1}^I\over \partial y }\right)
+{1\over 2}{{\partial}^2 {{\Sm}_1}^R\over {\partial y^2}} + 6x^2y^2
\tea
\noindent and
\bea
- {\partial{{\Sm}_1}^R\over \partial t}
&=& - \mu (x -g)
\left({\partial{{\Sm}_1}^I\over \partial x }\right)
+ \dot{g}
\left({\partial{{\Sm}_1}^R\over \partial x }\right)
+{1\over 2}{{\partial}^2 {{\Sm}_1}^I\over {\partial x^2}}\nn \\
& & - m (y - f)
\left({\partial{{\Sm}_1}^I\over \partial y }\right)
+ \dot{f}
\left({\partial{{\Sm}_1}^R\over \partial y }\right)
+{1\over 2}{{\partial}^2 {{\Sm}_1}^I\over {\partial y^2}} ,
\tea
where $f = y_0\cos {m t}$ .

\noindent  We then specialize the ansatz to the following
\bea
{{\Sm}_1 }^R &=& {\tilde \alpha}x^4 + {\tilde \beta}x^3 +{\tilde
\gamma}x^2 + {\tilde \sigma}x + \delta x^2 y^2 + {\tilde \omega}\nn \\
& & + \epsilon x y^2 + \kappa y^2 + \nu x^2 y + \theta xy + \xi y
\tea
and
\bea
{{\Sm}_1 }^I &=& {\tilde B}x^3 + {\tilde C}x^2 + {\tilde D}x +
{\tilde E} + F x^2 y \nn \\
& &+G x y^2 + L y^2 + M xy + Ny .
\tea
The explicit solutions for these quantities  are
written down in gory detail in
Appendix A for the sake of completeness.
However, we are only interested in comparing the
two solutions under the conditions that the initial conditions are
such that the microsuperspace and minisuperspace solutions start out as
close as possible. Therefore here we will present the solution to the
full problem under these conditions, with  a large number
of integration constants set to zero. In fact, we will set $y_0 = 0$
as well, which means the lowest order solution will be initially
centered around $y = 0$, which is necessary to have the solutions
to the lowest order start out as close as possible. Thus
in this case we have,
\bea
{\tilde \alpha} & =& \alpha ,\nn \\
{\tilde \beta}  & = & \beta ,\nn \\
{\tilde \gamma} & = & \gamma + {3\over 2\mu(\mu + m)} ,\nn \\
{\tilde \sigma} & = & \sigma + {3x_0\over m}t \sin \mu t ,\nn \\
\delta          & = & {3\over \mu + m} ,\nn \\
\epsilon        & = & {3\mu x_0\over m(m +\mu)}\cos \mu t , \nn \\
\kappa          & = & {3\over 2m(\mu + m )} - {3{\mu}^2{x_0}^2\over
2m({\mu}^2 - m^2)} \cos 2 \mu t ,
\tea
and $\nu , \theta$ and $\xi$ vanish.
We have plotted the $|\Psi_{micro}|^2$ and
$|\Psi_{mini}|^2\Big|_{y=0}$  for a succession of time intervals up
$t= 2\pi /\mu$, after which the secular terms
become dominant and the results
cannot be trusted. The wave functions in this section are
presented in their unnormalized form. It can be easily shown that
they are normalizable to $O({\lambda \over L})$. However, in the plots in

Figs. (1), (2), and (3) we have used a rough normalization of dividing
the probability density by the $x$ independent factor in the
unnormalized wave function. This seems to serve our purpose, since
we are more interested in tracking the motion of the center of the
wave packet. One can in principle numerically calculate a
normalization factor, but since such a
normalization can only be valid to order ${\lambda \over L}$, and, of course,
cannot be trusted beyond the timescale where the
secular terms start to dominate, we decided against this.
Notice that the behavior of the wave packets in the microsuperspace
and minisuperspace cases is at least qualitatively similar, in that
both are Gaussian-like packets that oscillate about $x = 0$ at roughly the
same frequency.
These results will be used in the next section to try to develop
a set of criteria for a ``good" microsuperspace.


\setcounter{equation}{0}
\section { Criteria for a Good Microsuperspace }
As we showed in the previous section, for  Case I microsuperspaces
the minisuperspace solution which begins as a Gaussian-like peak
around $y = 0$ and some value $x_0$ of $x$, and whose velocity is
initially zero, follows essentially the trajectory of the
microsuperspace (where $y$ and $p_y$ are set equal to zero before
quantization).  We have allowed our packet to evolve up to a point
where the secular terms begin to grow too large and the shape and
motion of the packet can no longer be trusted.

It is obvious in this case that the microsuperspace gives us at least
a qualitative picture of the behavior of the full minisuperspace wave
packet.  However, at least technically, the form and motion of the $y
= 0$ slice of the full packet is not the same as that of the
microsuperspace packet.  We can solve for the position of the peak of
$\mid \Psi \mid^2$ of the two packets by solving $\partial {{\cal
S}}^R/
\partial x = 0 $ in the microsuperspace case and $\partial {\cal S}^R/\partial
x \mid_{y=0} = 0$ in the minisuperspace case.  It is only reasonable
to solve these equations to first order in $\lambda /L$ because
${\cal S}$
is only valid to this order.  The peak of $\mid \Psi \mid^2$ will be
at $x_m = x_0 \cos \mu t + (\lambda /L) x_{1m}^{(0)}$ for the
microsuperspace and $x_m = x_0 \cos \mu t + (\lambda /L)
x_{1m}^{(1)}$ for the minisuperspace.  The $x_{1m}$ are
\bea
 x^{(0)}_{1m} &=& {{1}\over {\mu}}[ -{{x_0^3}\over {8\mu}}\cos 3\mu t
+ \left( {{15}\over {8\mu}} x_0^3 + {{3x_0}\over {2\mu^2}} \right )
\cos \mu t  + \left ( {{3x_0^2}\over {2}} + {{3x_0}\over {\mu}} \right
)t \sin
\mu t ],  \\
x^{(1)}_{1m} &=& x^{(0)}_{1m} + {{3x_0}\over {\mu (\mu + m)}} \cos
\mu t + {{3x_0}\over {m}} t \sin \mu t.
\tea
The changes in (4.2) are a slight shift in the zero-order amplitude,
$x_0 \cos \mu t$, and a change in the secular term $t \sin \mu t$.
Notice that for large $m$ (high mode number $n$ or small $L$) that
these corrections will be small.
As we saw from the graphs in Sec 3, these changes do not affect the
qualitative behavior of the wave packet, but it is obvious that they
do change the detailed behavior of the system.  Whether one says that
the microsuperspace system is a good model of the minisuperspace
system depends on exactly what questions one asks about the system.
However, given the coincidence of the graphs of the motion of the wave
packet in Sec. 3, we can reasonably say that
for Case I the microsuperspace is
``good'' in that it adequately represents the qualitative
behavior of the larger
minisuperspace.

For the other cases we will have to discuss in more detail the
meaning of a good minisuperspace.  In Case I the potential term in
$xy$-space has the form of a bowl (Figure 4),
and it is not surprising that
a wave packet exists that moves along the $y = 0$ line and is
confined to that line by the rising potential on either side of it.
The bowl shape also means that if the wave packet is given a small
initial motion in the $y$-direction it will (as can be seen in
Appendix A) only oscillate around the $y =0$ line while
following essentially the trajectory given in (4.2).  This leads us
to another question about the meaning of minisuperspace quantization.
We have based our ideas up to this point on the question of whether a
wave packet follows the minisuperspace (in our case microsuperspace)
trajectory when it ``feels'' the surrounding areas of superspace. The
new question is whether the trajectory of the wave packet is stable
with respect to small changes in the initial conditions that set it
up.  It is not entirely clear that such instability is a drawback,
since in the case of relativistic cosmology the initial conditions
are not ours to decide. However, too much dependence on initial conditions
to make a minisuperspace state a good description of the quantum
physics of the system is probably best avoided.  Of course, we cannot
escape all
dependence on initial conditions.  There are
examples, such as Bianchi type I cosmological minisuperspaces, where
the three-curvature of $t =$ const. slices that serves as a potential
term is zero, which means that trajectories are marginally stable and
any small changes in initial conditions in a wave packet will cause
the packet to move slowly away from any given minisuperspace
trajectory.  However, it is probably best to be suspicious of all
cases where the minisuperspace trajectory is, in the classical
regime, unstable with respect to small changes in initial conditions.
We will call such cases ``classically unstable'', and also use
``classically stable'' in the obvious way. A gravitational
minisuperspace example of a system that is unstable with respect to
slight changes in initial conditions is given in Ref. \cite{GRR2}.

Given these caveats we can now look at the cases other than Case I of
the $\lambda \phi ^4$ theory.  Case II ($\mu^2 < 0$, $\lambda >0$)
  corresponds to a
one-dimensional form of the usual ``Mexican hat'' potential of
inflationary theory.  If we consider the minisuperspace we must
divide Case II into two subcases, II$_a$ where $\mu^2 < 0$, $\lambda
> 0$, $m^2 > 0$, and II$_b$ where $\mu^2 < 0$, $\lambda > 0$, $m^2 <
0$. Of course, $m^2$ is greater or less than zero if $-\mid \mu
\mid^2 + (2n\pi /L)^2$ is greater or less than zero.
For II$_a$ Fig. (5a) shows the potential in $x$ and $y$.  Here the
$y=0$ line is always a minimum with respect to $y$ of the potential.
This means that the wave packet will be confined by the rising
potential
in the $y$-direction, and independent of the complicated motion in the
$x$-direction, we can expect that the packet will have the same
stable behavior as in Case I, so again II$_a$ will imply a ``good''
microsuperspace.  Case II$_b$ is less straightforward.  Figure (5b)
shows the potential there.  For small $x$ and $y$ the potential term
is similar to the ``crown'' of the Mexican hat potential, falling off
in all directions from $x = y = 0$.  As we move away from this point
along the $y = 0$ line we arrive at minima at $x = \pm (\mid \mu
\mid)/2)
\sqrt{L/\lambda}$, then the potential rises sharply as the $(\lambda
/L) x^4$ term takes over.  In the $y$-direction the $y = 0$ line is a
maximum of the potential at $x = 0$, and remains so until $\mid x\mid$
reaches
$(1/\sqrt{3} ) [1 - (2n\pi/\mid \mu \mid L )^2]^{1/2} (\mid \mu \mid /2)
\sqrt
{L/\lambda}$. So, between this value of $\mid x \mid$ and
$\mid x \mid  = (\mid \mu \mid /2)
\sqrt{L/\lambda}$ the potential in the $y$-direction has a relative
minimum at $y = 0$.  This implies that for Case II$_b$, small
oscillations of the wave packet around the $x$-minima have almost
exactly the same character as similar oscillations in Case I, so for
initial conditions that imply these oscillations, the microsuperspace
does give a reasonable approximation to the motion of the full
minisuperspace.  However, wave packets that begin near $x = 0$, or
those which oscillate around the $\mid x \mid = (\mid \mu \mid /2)
\sqrt {L/\lambda}$ minima whose
amplitudes of oscillation around these
minima are large enough for the wave packet to pass
the $\pm (1/\sqrt{3} ) [1- (2n\pi/\mid \mu \mid L)^2 ]^{1/2}
(\mid \mu \mid /2)
\sqrt{L/\lambda}$ points will behave differently.  As an example we
will consider a wave packet that begins near $x = 0$ and is as close
to $y = 0$, $p_y = 0$ as possible.  For small $x$ and $y$ the
potential is close to an inverted harmonic oscillator, and is
approximately $-{{1}\over {2}} \mu^2 x^2 - {{1}\over {2}} m^2 y^2$.
In this case there exists a wave packet that stays centered around $y
= 0$ while moving toward positive or negative $x$.  Such a wave
packet is discussed in Appendix B.  Here we
encounter the problem discussed above.  If we assume that the
existence of a packet centered around $y = 0$ that remains centered
around $y = 0$ during the evolution of the packet in $x$ implies
that the quantum microsuperspace $y = 0$ describes approximately the
evolution of the minisuperspace packet, then we must say that this
case allows a good microsuperspace.  If, however, we insist that the
microsuperspace must be stable against small changes in initial
conditions, it is obvious that giving the packet a small initial
momentum in the $y$-direction, will cause it to depart wildly from
the microsuperspace, and the $y = 0$ microsuperspace does not give a
good approximate description of the system.

For Case III($\mu^2 > 0, \lambda < 0$),
 $\mu^2 > 0$ implies that $m^2 > 0$.  The $y = 0$
potential has the form of a depression around $x = 0$ that rises to
maxima at $x = \pm (\mu /2) \sqrt{L/\mid \lambda \mid}$, and then
falls off rapidly, allowing quantum tunneling from the depression
into regions of larger $\mid x \mid$.  The potential is shown in
Figs. (6a,b).
Since $m^2 > 0$, for small $x$
the potential rises in the $y$-direction which means that in a small
range of $x$ near $x = 0$ wave packets are confined in $y$, and the
microsuperspace gives an adequate description of the minisuperspace
behavior.  However, as $x$ reaches the values $\pm (1/\sqrt{3}) [ 1 +
(2n\pi /\mu L)^2 ]^{1/2} (\mu / 2) \sqrt {L/ \mid \lambda \mid}$, the
potential goes from having a minimum at $y = 0$ to having a maximum
there.  Notice that this value of $x$ can be smaller or larger than
the maxima of the microsuperspace potential, $x = \pm (\mu /2)
\sqrt{L/\mid \lambda \mid}$(the two possibilities are shown in Figs.
[6a] and [6b], respectively).
If the value lies inside the maxima,
then a wave packet that oscillates around $x = 0$
can reach a region
where the $y = 0$ line is classically unstable if the amplitude of
its oscillation is large enough.  If the value of $x$ where the $y =
0$ line becomes a maximum is outside the maxima of the
microsuperspace potential, then if the amplitude of oscillation of
the wave packet is large enough, it will escape from the depression
and will begin to roll down the slope toward regions of $\mid x \mid
> (1/\sqrt{3}) [ 1 + (2n \pi /\mu L)^2 ]^{1/2} (\mu /2) \sqrt{L /\mid
\lambda \mid}$ into classically unstable regions of the $y = 0$ line.
Notice that in all cases there exists the possibility of quantum
tunneling from the depression out into regions of classical
instability or areas of $x$ where it is possible to roll down into
such regions.  The possibility of tunneling means that if classical
stability is a criterion for the microsuperspace to give an adequate
description of the quantum dynamics of the full system, Case III can
never be considered a good microsuperspace.  The only exception is if
we call the microsuperspace ``good'' if it gives a reasonable
description of the minisuperspace system for a time, and then breaks
down.  In this case the tunneling time could be long, and the wave
packet could stay in the depression for a long enough time to be
useful for the predictions one wants to make.

As in Case II, Case IV ($\mu^2 < 0$, $\lambda < 0$)
gives rise to two sub-cases in the
minisuperspace, Case IV$_a$ where $\mu^2 < 0$, $\lambda < 0$, $m^2
>0$, and Case IV$_b$ where $\mu^2 < 0$, $\lambda < 0$, $m^2 < 0$.  In
both sub-cases, which are shown in Figs. (7a,b), near $x = 0$
the microsuperspace potential can be modeled by that of an
upside-down harmonic oscillator, and the wave packet given by Eq.
(B.12)
will in general roll off $x = 0$ in the direction of larger $\mid x
\mid$.  In the minisuperspace in Case IV$_a$ near $x = 0$ the $y = 0$
line is a minimum, so a $y = 0$ trajectory is classically stable.
However, when $x = \pm (1\sqrt{3}) [(2n\pi/
\mid \mu \mid L)^2 - 1
]^{1/2} (\mid \mu \mid /2 ) \sqrt {L/\mid \lambda \mid}$
the $y = 0$ line
becomes a maximum in $y$, and there is nothing to prevent the wave
packet from moving to these values of $x$, so it will always reach
regions of classical instability.  In Case IV$_b$ near $x = y = 0$
we can always construct a wave packet along the lines of (B.12) that will
stay centered around $y = 0$ while moving along that line.  However,
every point on the $y = 0$ line is classically unstable, so, again, if
classical instability is a criterion for the usefulness of the
microsuperspace, then in Case IV$_b$ it must be assumed that the
microsuperspace description does not give a reasonable approximation
to the full minisuperspace behavior.

In the next section we will try to summarize the status of
the problem of when the
$y = 0$ microsuperspace provides a useful description of the full
minisuperspace behavior and compare the criteria we have developed to
possible gravitational scenarios.
\setcounter{equation}{0}
\section{ Conclusion and Discussion}
We can summarize our conclusions from Section 4 as follows.
The full potential term $({{\mu}^2\over 2})x^2 + \left({\lambda\over
L}\right)x^4 + {m^2\over 2} y^2 + 6 \left({\lambda\over L }\right)
x^2 y^2$ determines whether the microsuperspace gives a good
approximation to the behavior of a minisuperspace wave packet
centered around the $y=0$ sector. In Ref.\cite{Kuchar2}
a gravitational example
of a case where the minisuperspace wave packet had behavior that
diverged wildly from the microsuperspace behavior was given. In all
of the cases of Section 4 a state that remained peaked around
the $y=0$ minisuperspace during its whole evolution in the
$x$-direction existed. The motion of the peak of the minisuperspace
packet did show a minor deviation from the motion of the
microsuperspace packet, so one would argue that none of the
microsuperspaces were truly ``good"
in the sense of faithfully representing the detailed
quantitative behavior of the minisuperspace. However, the microsuperspace
and minisuperspace behaviors were qualitatively similar so that
it is probably best to say that all of the microsuperspaces gave
a reasonable approximation to the true minisuperspace behavior.

The one major difference was in the stability of the packets against
small changes of the $y$-position and $y$-velocity, which may be
taken to be a criterion for the ``badness" of the microsuperspace. If
one does so, then in cases II$_b$ , III, and IV the
microsuperspace cannot be taken to represent the behavior of the
full minisuperspace.

As we stated in the Introduction, here we have considered an
approach different from Ref.\cite{Kuchar}  (henceforth referred to as KR1)
to define a good microsuperspace. However, since we have used
exactly the same model, it is worth making a few comments comparing
the two approaches. In KR1, the idea was so start with the full
Schr\"odinger equation for superspace (minisuperspace in our
language) , expand the wave function in eigenstates of the
minisuperspace Hamiltonian ( the part of the full Hamiltonian that
depends on $y$ and $p_y$) that parametrically depend on $x$, and
find  the conditions under which the full Schr\"odinger equation
could be reduced to a ``projected Schr\"odinger equation " for the
set of ``microsuperspace wave functions" ( the set of
$x, t$ dependent coefficients in the above stated mode expansion)
evolved by the pure microsuperspace  Hamiltonian ($y=0, p_y
= 0$). Expectation values of the microsuperspace operators could
then be obtained  from the density matrix  constructed from the
wave functions.
The criteria that emerge are roughly akin to the Born-Oppenheimer
approximation, i.e, i)the parametrized eigenstates should vary
slowly with $x$ and ii)the $x$ dependent energy eigenvalues should
be small compared with the microsuperspace potential.
There was no attempt to consider different signs of $\lambda$
and $\mu$, so the KR1 model is essentially equivalent to only
Case I.

In our analysis, rather than trying to reduce the Schr\"odinger
equation on the larger space , we compare the two behaviors
on the level of solutions. Our minisuperspace wave function is
a coherent-state-like solution of the full Schr\"odinger equation
while the microsuperspace wave function is a similar solution
to the microsuperspace Schr\"odinger equation $(p_y = 0, y = 0)$,
which has {\it no} information on the $y$ degree of freedom.
So we see that the approaches differ slightly on what is meant
by the microsuperspace wave function. The differences also lie
in that ours is formulated in terms of ``coherent" states for
the full solution, while KR1 relies on a decomposition in terms of
``energy" eigenstates and the criteria are stated in terms of
excitation levels. However, the approaches can be related by
writing the wave function (3.16) in terms of the eigenstates of KR1,
and testing the slow variation criterion, for example. Though the two notions
of the microsuperspace wave function do not quite coincide, we believe
that criterion ii) can be closely related to the criterion that the
difference in the trajectories of the centers of the
$y=0$ minisuperspace wavepacket
and the microsuperspace wavepacket should be small. Our classical
stability
criterion does not seem to have an obvious parallel in KR1.
Overall, it is probably fair to say that while our approach may
involve some loss of generality as opposed to KR1 since it relies
on a specific type of solution, it has the advantage that it
allows one to follow the evolution in greater detail and formulate
more concrete criteria.

The model of Ref.\cite{Sinhu} differed from the
one considered here in that it considered an infinite-dimensional
minisuperspace that was coupled to a curved rather than flat space
and the treatment was as in KR1 based on the analysis of equations
of motion rather than solutions. Apart from the features specifically
tied to curved spacetime, the requirement there of the microsuperspace
potential dominating the backreaction term is loosely analogous to
the criterion of the centers of the two wave packets remaining close.
However, an extension of the treatment of this paper to include
all the infinite number of modes as in \cite{Sinhu} would be a worthwhile
exercise. Of all of the previous attempts at finding criteria that
tell us whether a particular quantum minisuperspace is a good
approximation to a real solution, the present article is most
closely related to \cite{Kuchar2}. There, as here, the motion of a
microsuperspace wave packet was compared directly to that of a
minisuperspace wave packet. There, however, as we have just
mentioned, the behavior of the two packets was extremely divergent,
while here they are qualitatively similar.

Before going on to the case of gravity, we would like to mention
another possible approach that can be tested on quantum mechanical
and field theoretical models. It is clear that if we only
want to know expectation value of operators that are constructed
solely from minisuperspace variables, given the full wave function
, we only require the knowledge of the reduced density matrix
(${\rho}_{red} = Tr_{y} |\Psi><\Psi| )$ to calculate
these objects. A similar idea was discussed in KR1, though it
required more specific assumptions about the expansion of
the wave function in energy eigenstates. The evolution of the
reduced density matrix would be guided by a master equation rather
than a Hamiltonian evolution through the Schr\"odinger equation.
The master equation will contain diffusion and dissipation
terms arising
from the averaged effect of the minisuperspace modes,
 in addition to the pure microsuperspace Liouville operator.
Demanding the smallness of these extra terms will then lead to
criteria for the goodness of the microsuperspace. This approach can
be thought of as elevating the approach of \cite{Sinhu} to the quantum
statistical mechanics level.

The next important question is of course to understand how one
can extend the analysis of this paper to possible gravitational
scenarios. As mentioned in the Introduction, the dynamics in
this case will be dictated by a Wheeler-DeWitt equation (1.1)
rather than a Schr\"odinger equation and we will immediately
be burdened with the host of problems that have posed a barrier
to the construction of a successful theory of quantum gravity
until now. However, without trying to confront the full theory
in all its complexity, one can conceive of  gravitational
minisuperspace models
(not necessarily finite-dimensional) which contain microsuperspaces
embedded in them that are exactly or approximately solvable, as
for example in \cite{Kuchar2}, and apply a similar analysis to them . Two
appropriate candidates appear to be the Gowdy model \cite{Gowdy}
, which has a Bianchi- I microsuperspace embedded
in it, and Halliwell and Hawking's model\cite{H-Hawk} of a closed Robertson
Walk
   er
universe with gravitational perturbations.

In such a picture, as stated before, $gR$  is the appropriate
analog of our potential function which is a determining factor
in the goodness of the microsuperspace. Provided one can define an
appropriate ``time" in such a model, it is clear that while
parameters analogous to $\lambda$ and $\mu$ are fixed by the
given gravitational Hamiltonian (as opposed to being free
parameters that we can vary at will as in this paper), they are
necessarily time dependent. The major and nontrivial task would
then be to understand how to construct criteria analogous to ours for
time dependent coefficients.
Another point that perhaps is worth mentioning is that though
$gR$ is possibly the major factor in deciding on the criteria, the
superspace metric $G_{ijkl}$ will also surely play a role. It
so happens that all the gravitational models that have been analyzed
in this context \cite{Kuchar2} possess a flat superspace metric. It is
therefore important to analyze models that have a curved superspace
metric to study this particular feature.

An interesting point is that it may be possible to formulate
stability criteria for microsuperspace trajectories
in rather general terms without referring to a specific model.
A solution to the Einstein equations is represented by a
driven geodesic in superspace which obeys the equation (for $g_{00} =
-1$, $g_{0i} = 0$)
\be
{{d^2 g^{(ij)}}\over {ds^2}} + \Gamma^{(ij)}_{(k\ell )(mn)}
{{dg^{(k\ell )}}\over {ds}} {{dg^{(mn)}}\over {ds}} = {{\delta (\sqrt{g}
R )}\over {\delta g_{(ij)}}},
\te
where the $\Gamma^{(ij)}_{(k\ell )(mn)}$ are the superspace
Christoffel symbols constructed from $G_{ijk\ell}$,
and $\sqrt{g}R$ is defined above.
Classical stability
of a trajectory in superspace can be studied by considering the
appropriate superspace ``equation of geodesic deviation" for the
trajectory. This type of equation can be used for at least
two purposes. It is possible to find regions of ``classical
instability " of the type found in cases II, III and IV.
It will also serve, by finding examples of classical ``magnetic
mirror" behavior such as those that exist in type IX models, as
an indicator of the existence of quantum ``magnetic mirror"
solutions such as those found in Ref.\cite{Kuchar2}

\vspace{2 cm}
\noindent{\bf Acknowledgement} \\
The research reported here was supported in part by an NSF-CONACYT
grant between the Instituto de Ciencias Nucleares, UNAM and the
University of Utah (P-89).

\renewcommand{\thesection}{A}
\renewcommand{\theequation}{A.\arabic{equation}}
\setcounter{equation}{0}

\section{Appendix A}
Let us first treat the microsuperspace sector.
Substituting the ansatz (3.14) in equations (3.12) and (3.13)
and equating equal
powers of $x$, we obtain the following coupled equations for the
unknown coefficients. For the imaginary part we get
\bea
\dot{B} &=& g - 3\mu \beta ,\\
\dot{C} &=& 3 \mu g \beta - 2\mu \gamma - 3 B\dot{g} + {3\over
2\mu} ,\\
\dot{D} &=& 2\mu g \gamma - \sigma \mu - 2\dot{g}C + 3\beta ,\\
\dot{E} &=& \mu g \sigma - D\dot{g} + \gamma ,
\tea
and for the real part,
\bea
-\dot{\beta} &=& -3 \mu B + {\dot{g}\over \mu} ,\\
-\dot{\gamma} &=& 3 \mu g B - 2\mu C + 3\dot{g}\beta ,\\
-\dot{\sigma} &=& 2 \mu g C - \mu D + 2\dot{g}\gamma +3B ,\\
-\dot{\omega} &=& \mu g D +\dot{g}\sigma + C ,
\tea
and $\alpha = {1\over 4\mu}$.
These equations can be decoupled in the following manner. From (A.5)
one can solve for $B$ in terms of $g$ and $\beta$ and substitute it
into (A.1) to obtain the following differential equation for $\beta$.
\be
\ddot{\beta} + 9{\mu}^2 \beta = 4\mu x_0 \cos{\mu t}.
\te
which is simply an equation for a harmonic oscillator driven by a
known oscillating source, and can be easily solved.
The solution can be written down as
\be
\beta = {x_1}^3\cos{3\mu t} + {x_0\over 2\mu}\cos{\mu t} ,
\te
where $x_1$ is an arbitrary constant. We follow the same procedure
for the rest of the functions.

\noindent Solving for $C$ from (A.6) , substituting in (A.2) and using
the now known solution for $\beta$ we get the following equation for
$\gamma$
\be
\ddot{\gamma} + 4 {\mu}^2 \gamma = 18{\mu}^2 x_0 {x_1}^3\cos{4\mu t}
+ 6 \mu {x_0}^2\cos{2\mu t} + 3 .
\te
To obtain the equation for $\sigma$ one solves for $D$ from (A.7) and
substitutes this in eqn. (A.3). Using the known solutions for $\beta$
and $\gamma$ one obtains the following equation
\bea
\ddot{\sigma} + {\mu}^2\sigma &=& 9 \mu {x_1}^3 \cos{3\mu t}
-18{\mu}^2{x_0}^2{x_1}^3\cos{5 \mu t} \nn \\ &+& 8 {\mu}^2x_0 C_1 \cos{3\mu
t}
+ 12 {\mu}^2 {x_0}^3 t \sin{3 \mu t} + 6 x_0\cos{\mu t} .
\tea
As before, both (A.11 ) and (A.12) are equations for driven harmonic
oscillators and their solutions , which are explicitly written down
in eqn. (3.15) can be obtained by standard methods. The solution
for $\omega$ is trivial, since it can be reduced to quadratures, and
it is easy to see that the right hand side of (A.8) can be written
as a total derivative leading to the solution for $\omega$ given in
(3.15). One can obtain $B, C, D, E$  easily as well, writing them in
terms of the coefficients of the real parts. However, as stated
before in the text, we will not calculate them since they are
unimportant for our considerations.

Now let us treat the minisuperspace sector. Substituting (3.20) and
(3.21) in eqns. (3.18 ) and (3.19) and
equating equal powers of $x$ and $y$, we
obtain the following set of equations for the coefficients of the
imaginary part
\bea
\dot{\tilde{B}} &=& - 3\tilde{\beta}\mu + 4{\tilde{\alpha}} \mu g , \\
\dot{\tilde{C}} &=& -2\tilde{\gamma}\mu + 3\tilde{\beta}\mu g - 3
\tilde{B}\dot{g}
+ 6\tilde{\alpha} + \nu m f - \dot{ f} F + \delta ,\\
\dot{\tilde{D}} &=& -\tilde{\sigma} \mu + 2\mu\tilde{\gamma} g
- 2\dot{g}\tilde{C}
+ 3{\tilde{\beta}} + m f \theta - M\dot{f} + \epsilon ,\\
\dot{\tilde{E}} &=&
 \sigma \mu g - \dot{g}\tilde{D} + \gamma + mf\xi
-N\dot{f} + \kappa ,\\
\dot{F} &=& -(2 \mu + m )\nu + 2mf\delta , \\
\dot{G} &=&  -(\mu + 2m)\epsilon + 2\mu g \delta , \\
\dot{L} &=& \mu g \epsilon - \dot{g}G + \delta - 2m\kappa ,\\
\dot{M} &=& -  (\mu + m)\theta + 2 \mu g \nu - 2F\dot{g}
+2\epsilon m f - 2\dot{f}G , \\
\dot{N} &=& \theta \mu g - \dot{g}M + \nu - m \xi + 2mf\kappa -
2\dot{f}L .
\tea

For the real part we get
\bea
-\dot{\tilde{\beta}} &=& -3\tilde{B}\mu + 4\dot{g}\tilde{\alpha} ,\\
-\dot{\tilde{\gamma}} &=& -2\mu \tilde{C} + 3\tilde{B}\mu g +
3\dot{g} \tilde{\beta} + mfF + \dot{f}\nu ,\\
-\dot{\tilde{\sigma}} &=& -\tilde{D}\mu + 2\mu g \tilde{C} +
2\tilde{\gamma}\dot{g} +3\tilde{B} + mfM + \theta\dot{f} + G ,\\
- \dot{\tilde{\omega}} &=& D\mu g + \dot{g}\tilde{\sigma} + \tilde{C}
+\dot{f}\xi + L + mfN ,\\
-\dot{\epsilon} &=& -(\mu + 2m)G + 2\delta\dot{g} ,\\
-\dot{\nu}      &=& -(2\mu + m)F + 2\delta\dot{f} , \\
-\dot{\theta} &=& -M\mu + 2\mu g F + 2\dot{g}\nu - Mm + 2mfG +
2\dot{f}\epsilon ,\\
-\dot{\kappa} &=& \mu gG +\dot{g}\epsilon - 2Lm ,\\
-\dot{\xi}     &=& \mu g M + \dot{g}\theta + F - Nm + 2mfL +
2\kappa\dot{f},
\tea
where $\tilde{\alpha} = \alpha ={1\over 4\mu}$ and $\delta = {3\over
\mu + m}$. We will use the same technique as in the microsuperspace
case to decouple these equations. As before, we solve for an
imaginary-part coefficient in terms of the real-part coefficient and
substitute in the corresponding equation for the time derivative of
the real-part coefficient to obtain a differential equation for it.
We see that the equations for $\tilde{\beta}$ and $\tilde{B}$ are
identical to their microsuperspace counterparts, so that we
conclude $\tilde{\beta} = \beta$. From inspection of the equations
(A.13)
-- (A.30), we see that the pairs that should be used in the above
solve and substitute procedure are the following : $[F,\nu] , [C, \gamma],
[G,\epsilon], [M, \theta], [D,\sigma], [L, \kappa]$ and $[N ,\xi]$.
The process of solving  also proceeds in the above order, so
the solution of a preceding variable can be used to solve for
the subsequent one as before.

\noindent Using the above procedure, the equation satisfied by $\nu$
is given by
\be
\ddot{\nu} + {(2\mu + m)}^2 \nu = 12 m y_0 \cos{m t}.
\te
In all of what follows, we will write down the solution to the forced
harmonic oscillator equations as only the part coming from the
inhomogeneous term, using the freedom in the arbitrary coefficients
to set the homogeneous solution to zero. This is done only for
simplification and it is obvious from the procedure how to
incorporate the homogeneous solution. Bearing this in mind, the
solution to (A.31) can be written as
\be
\nu = {3my_0\over \mu(\mu +m)}\cos{m t}.
\te
Next, the equation for ${\tilde{\gamma}}$ is given by
\be
\ddot{\tilde{\gamma}} = 18{\mu}^2 x_0 {x_1}^3\cos{4 \mu t}
+ 6\mu {x_0}^2\cos{2 \mu t} + 6{m^2 {y_0}^2\over \mu}\cos{2 mt}
+ {6\mu\over m +\mu} +3  ,
\te
and the solution is
\be
\tilde{\gamma} = {3\over 4 {\mu}^2}\left( {3\mu +m\over \mu +
m}\right) -{3\over 2}{x_0}^2{x_1}^3\cos{4\mu t} +{3\over 2} {x_0}^2
t\sin{2\mu t} + {3\over 2}{m^2 {y_0}^2\over \mu({\mu}^2 - m^2)}\cos{2
mt}  .
\te
The equation for $\epsilon$ is given by
\be
\ddot{\epsilon} + {(\mu + 2m)}^2\epsilon = 12\mu x_0 \cos{\mu t} ,
\te
and the solution is
\be
\epsilon = {3\mu x_0\over m(m +\mu)}\cos{\mu t} .
\te
The equation for $\theta$ is given by
\be
\ddot{\theta} + {(\mu + m)}^2\theta = 12(m + \mu)x_0 y_0\cos{(\mu +
m)t} ,
\te
and the solution is
\be
\theta = 6x_0 y_0 t \sin{(\mu + m)t}.
\te
The equation for $\kappa$ is given by
\be
\ddot{\kappa} + 4m^2 \kappa = {6m\over \mu + m} +
{6{\mu}^2{x_0}^2\over m} \cos{2\mu t} ,
\te
and the solution is
\be
\kappa = {3\over 2m(\mu + m)} - {3{\mu}^2{x_0}^2\over 2m ({\mu}^2 -
m^2)} \cos{2\mu t} .
\te
The equation for $\tilde{\sigma}$ is given by
\bea
\ddot{\tilde{\sigma}} + {\mu}^2 \tilde{\sigma} &=&
6x_0 \left (1 + {\mu\over m} \right) +
3\mu (4{x_1}^3 - {x_0}^3)\cos {3\mu t}
 + 12 {\mu}^2{x_0}^3 t \sin{3
\mu t} \nn \\
& - & 18{\mu}^2 {x_0}^3{x_1}^3\cos{5 \mu t}
+ {6 {x_0}{y_0}^2 m (2m - \mu)\over \mu - m}\cos (\mu + 2m)t\nn \\
&+& 12{x_0}{y_0}^2 m (\mu + m) t \sin (\mu + 2m)t ,
\tea
and the solution is
\bea
\tilde{\sigma} &=& -{3\over 2\mu}\left({x_1}^3 + {1\over
2}{x_0}^3\right)\cos{3 \mu t} + {3\over 4}{x_0}^2{x_1}^3\cos{5 \mu t}
-{3{x_0}{y_0}^2\mu\over 2m({\mu}^2 -m^2)}\cos(\mu + 2m)t \nn \\
&+& 3x_0 {(m + \mu)\over m\mu}t\sin{\mu t} - {3\over
2}{x_0}^3t\sin{3\mu t} - 3{x_0}{y_0}^2 t \sin (\mu + 2m)t .
\tea
and finally, the equation for $\xi$ is given by
\bea
\ddot{\xi} + m^2 \xi &=& 6 {y_0}\left[{m(m + 3\mu)\over \mu(\mu +
m)}\right]\cos{mt}
- {6\mu {x_0}^2{y_0}\over (\mu - m)}(2\mu - m)\cos(2\mu +m)t \nn \\
&+& 12\mu{x_0}^2{y_0}(\mu + m) t \sin (m + 2 \mu)t ,
\tea
and the solution  is
\be
\xi = {3{x_0}^2{y_0}m^2\over 2\mu({\mu}^2 - m^2)}\cos(2\mu + m)t
+ {3y_0 (m + 3\mu)\over \mu (m + \mu)}t\sin {m t}
-3{x_0}^2y_0 t \sin(m + 2\mu)t .
\te
The solutions that appear in (3.23) are a special case of these with
$y_0 = 0$.  Note that a small $y_0$ implies (from the form of ${\cal
S}_0$) that the peak of the wave packet will oscillate with small amplitude
around the $y = 0$ line.
It can be shown that $\tilde{\omega}$ again can be obtained
from a total derivative, but we do not demonstrate this explicitly
since this acts merely as a normalization factor.

\renewcommand{\thesection}{B}
\renewcommand{\theequation}{B.\arabic{equation}}
\setcounter{equation}{0}

\section{Appendix B}
Here we will demonstrate the quantum solution to the zeroth order
microsuperspace solution for ${\mu}^2 < 0 $, which is essentially
equivalent to a one dimensional upside down harmonic oscillator.
The equations to be solved are again (3.5) and (3.6) with ${\mu}^2$
replaced by $- {|\mu|}^2$. We make the following ansatz for
${\cal S}$
\be
{\cal S} = \bar{\alpha}x^2 + \bar{\beta}x + \bar{\gamma}
+ i\bar{A}x^2 + i \bar{B} x  + i \bar{C} .
\te
Plugging in the above ansatz into the modified equations (3.5) and
(3.6) we obtain the following equations for the unknown coefficients
\bea
\dot{\bar{\alpha}} &=& 4{\bar{\alpha}}\bar{A} ,\\
\dot{\bar{\beta}}  &=& 2 (\bar{\alpha}\bar{B} + \bar{A}\bar{\beta}) ,
\\
\dot{\bar{\gamma}} &=& \bar{\beta}\bar{B} - \bar{A} ,\\
\dot{\bar{A}} &=& {1\over 2}( 4{\bar{A}}^2 - 4{\bar{\alpha}}^2 -
{|\mu|}^2) ,\\
\dot{\bar{B}} &=& 2(\bar{A}\bar{B} - \bar{\alpha} \bar{\beta} ) ,\\
\dot{\bar{C}} &=& {1\over 2}( {\bar{B}}^2 - {\bar{\beta}}^2 +
2\bar{\alpha}) .
\tea
{}From the above equations we obtain the following equation for
$\bar{A}$
\be
\ddot{\bar{A}} - 12\bar{A}{\dot{\bar{A}}} + 16 {\bar{A}}^3 -
4{|\mu|}^2 \bar{A} = 0 ,
\te
and thus from eqn. (B.2), $\bar{\alpha}$ can be written down as
\be
\bar{\alpha} = C_0 \exp\left[{4\int \bar{A} dt}\right]
\te
where $C_0$ is an arbitrary constant.
The equation for $\bar{B}$ is given by
\be
\ddot{\bar{B}} - 4\bar{A}\dot{\bar{B}} + (6{\bar{A}}^2 +
2{\bar{\alpha}}^2  - 2 \dot{\bar{A}})\bar{B} = 0 ,
\te
where the known solutions for $\bar{A}$ and $\bar{\alpha}$ are
supposed to be inserted where these quantities appear.
Then $\bar{\beta}$ can be determined from
\be
\bar{\beta} = {2\bar{A}\bar{B} - \dot{\bar{B}}\over 2\bar{\alpha}}
\te
using the known solutions of $\bar{A}, \bar{\alpha}$ and $\bar{B}$.
Subsequently, from eqns.(B.4) and (B.7) we see that $\bar{\gamma}$ and
$\bar{C}$ are reduced to quadratures. The normalized wavefunction
can be written as:
\be
\Psi(x,t) = {\left({2{\bar{\alpha}}\over \pi}\right)}^{1\over 4}
e^{-{\bar{\gamma}\over 2} -{{{\bar{\beta}}^2\over 4\bar{\alpha}}
-i\bar{C}}}
e^{- (\bar{\alpha} + i\bar{A})x^2 - (\bar{\beta} + i \bar{B})x }.
\te

Though eqn.(B.12) formally gives a complete solution to the problem,
in practice it is quite difficult to find an exact solution since
eqns. (B.8) and (B.10) are highly nonlinear.
However, a special exact solution
was given by Guth and Pi in Ref.\cite{GuPi}. In our notation this solution
corresponds to
$\bar{B} = \bar{\beta} = 0$ and
\bea
\bar{\alpha} &=& {|\mu|\sin{2\phi}\over 2\left(\cos{2\phi} +
\cosh(2|\mu|t) \right)} \\
\bar{A} &=& {-|\mu|\sinh(2|\mu|t)\over  2\left(\cos{2\phi} +
\cosh(2|\mu|t) \right)} ,
\tea
where $\phi$ is a real constant of integration related to the width
of the wave packet at $t=0$ such that the wave packet is at its
minimum width at $t=0$. $\bar{C}$ and $\bar{\gamma}$ can be
obtained from the complex identity:
\be
{\left[b\cos(\phi - i|\mu|t)\right]}^{-{1\over 2}}
= 2{\bar{\alpha}}^{-{1\over 4}} e^{-{\bar{\gamma}\over 2} - i\bar{C}}
,
\te
where $b = {\left(|\mu|\sin{2\phi}\right)}^{-{1\over 2}}$
This corresponds to a wave packet whose center remains at the top
of the potential hill at $x=0$ for all times while the spread
grows with time.

\noindent From the wave function (B.12) it can be seen that the center
of our wave packet lies at $x = {\bar{\beta}\over 2\bar{\alpha}} =
<x>$.
We can solve for the motion of the center of
the wave packet quite easily without having to
find the general solution for the wave function.
Using eqns. (B.2) -- (B.7), one can show that
${\bar{\beta}\over 2\bar{\alpha}}$ satisfies the following
equation
\be
{\left(   {\bar{\beta}\over 2\bar{\alpha}}\right)}^{\cdot \cdot}
- {|\mu|}^2 {\left(   {\bar{\beta}\over 2\bar{\alpha}}\right)}
= 0 ,
\te
which has the simple solution
\be
{\bar{\beta}\over 2\bar{\alpha}} = x_0 \cosh{|\mu|t}  + {p_0\over
|\mu|}\sinh{|\mu|t}.
\te
where $x_0$ , $p_0$ are  arbitrary constants which can be interpreted as
the initial position and momentum of the center of the wave packet
respectively.
Eqn. (B.16)
is also merely a statement of Ehrenfest's theorem for this problem.
{}From (B.17) it is clear that (B.13) and (B.14)
gives a special solution with $x_0 =
0$ and $p_0 = 0$,
and from (B.17) it is also evident that for any other initial conditions
the wave packet will rapidly roll down the potential hill.

The above exact solution can also be used to obtain a minisuperspace
solution , i.e, in the two dimensional ($x$- $y$) problem. In the
approximation that we consider a region around the origin such that
$x$ and $y$ are small, for $m^2 <0$,
the full Hamiltonian can be approximated by
one for two decoupled upside down harmonic oscillators since the
coupling terms are of higher than quadratic order. Going through the
the same steps as the above one-dimensional case, one then has a
minisuperspace solution that is a two-dimensional wave packet whose
center remains at $x = y =0$ for all times while its spread grows in
time. Thus the solution will be a product of two Gaussian wave
packets, one in $x$ and the other in $y$. The center of the wave
packet in $y$ will obey an equation identical to (B.16)  with $|\mu|$
replaced by $|m|$.


\newpage

\noindent {\bf FIGURE CAPTIONS}
\vskip 1 true cm
\noindent Figure 1. The motion of the microsuperspace wave packet.
We show the packet at four different times in Figs. 1a-1d as graphs
of $|\Psi_{micro}|^2$ divided by $e^{-2(\lambda /L)\omega}$ with
$\mu
= 1$, $x_1 = C_1 = d_1 = 0$, $\lambda /L = 0.1$,
and $x_0 = 1$.  The four plots of Figs.
1a-1d are for $t = 0$, $t = 3\pi /4$, $t = 5\pi /4$ and $t = 2\pi$
respectively.
\vskip 0.5 true cm
\noindent Figure 2.  The motion of the $y = 0$ slice of the
minisuperspace wave packet.  As in Fig. 1 we show the packet at four
different times in Figs. 2a-2d as graphs of
$e^{2(\lambda /L)\tilde \omega}|\Psi_{mini}|^2 \big
|_{y = 0}$ with $\mu =
1$, $m = 2$, $\lambda /L = 0.1$, $x_0 = 1$, $y_0 = 0$,
and $x_1 = C_1 = d_1 = 0$ as
before.  The four plots of Figs. 2a-2d are for $t = 0$, $t = 3\pi
/4$, $t = 5\pi /4$ and $t = 2\pi$ respectively.
\vskip 0.5 true cm
\noindent Figure 3.  Representative plots of
the same $e^{2(\lambda /L)\tilde \omega}|\Psi_{mini}|^2$
as in Fig. 2
for $t = 0$ in Fig. 3a and $t = 3\pi /4$ in Fig. 3b, but
shown in the $xy$-plane.
\vskip 0.5 true cm
\noindent Figure 4.  The potential $({{\mu^2}\over {2}})x^2 +
({{\lambda}\over {L}}) x^4 + ({{m^2}\over {2}}) y^2 +
6({{\lambda}\over {L}})x^2 y^2$ for $\mu^2 = 6$, $m^2 = 20$, and
$(\lambda /L) = 1/4$.  This potential corresponds to Case I. Notice
that here and in the figures that follow we have taken $\lambda /L$
larger than one should for a valid perturbation approximation
in order to show the details of the structure of the potentials.
\vskip 0.5 true cm
\noindent Figure 5.  Figures 5a and 5b show $({{\mu^2}\over {2}}) x^2
+ ({{\lambda}\over {L}}) x^4 + ({{m^2}\over {2}}) y^2 +
6({{\lambda}\over {L}})x^2 y^2$ for $\mu^2$ negative
and $\lambda /L = 1$.
Figure 5a corresponds to Case IIa and has $\mu^2 = -1$, $m^2 = 20$.
Figure 5b
corresponds to Case IIb and has $\mu^2 = -7$ and $m^2 = -6$.
\vskip 0.5 true cm
\noindent Figure 6.  Figures 6a and 6b show $({{\mu^2}\over {2}}) x^2
+ ({{\lambda}\over {L}}) x^4 + ({{m^2}\over {2}}) y^2 +
6({{\lambda}\over {L}})x^2 y^2$ for $\mu^2 = 2$, $\lambda /L = -1$
(Case III).  Figure 6a shows the potential for $m^2 = 3$ where the
points where the $y = 0$ line changes from a minimum to a maximum in
$y$ occur inside $x = \pm (\mu /2) \sqrt {L/|\lambda|}$.  In Figure
6b $m^2 = 7$, so the points where the $y = 0$ line changes from a
minimum to a maximum in $y$ occur outside $x = \pm (\mu /2) \sqrt{L
/|\lambda |}$.
\vskip 1 true cm
\noindent Figure 7.  Figures 7a and 7b show $({{\mu^2}\over {2}}) x^2
+ ({{\lambda}\over {L}}) x^4 + ({{m^2}\over {2}}) y^2 + 6({{\lambda}\over
{L}}) x^2 y^2$ for $\mu^2$ negative and $\lambda /L = -1$ (Case IV).  Figure
7a shows this potential for $\mu^2 = -1$ and $m^2 = 6$ (Case IVa),
where the $y = 0$
line is a minimum in $y$ until $x = \pm (1 /\sqrt {3})[(2n\pi / |\mu
| L) - 1]^{1/2} (|\mu |/2) \sqrt {L/ |\lambda |}$ where it becomes a
maximum in $y$.  Figure 7b corresponds to $\mu^2 = -7$, $\lambda /L =
-1$, and $m^2 = -6$ (Case IVb).  Here the potential falls off from $x
= y = 0$ in all directions.
\end{document}